\documentstyle[epsf,twocolumn,seceq]{jpsj}
\def\runtitle{{\sc LETTERS}}
\def\runauthor{{\sc LETTERS}}
\title{Orbital-Degenerate Paramagnetic Metal Sr$_{2}$MoO$_{4}$:\\
An Electronic Analogue to Sr$_{2}$RuO$_{4}$}
\author{Shin-Ichi {\sc Ikeda}, Naoki {\sc Shirakawa},
Hiroshi {\sc Bando} and Youiti {\sc Ootuka}$^{1}$}
\inst{Electrotechnical Laboratory, Tsukuba 305-8568\\
$^{1}$Department of Physics, Tsukuba University, Tsukuba 305-8577}
\recdate{July 3, 2000}
\abst{We present the first systematic study on polycrystalline Sr$_{2}$MoO$_{4}$ 
as an electronic analogue to the spin-triplet superconductor Sr$_{2}$RuO$_{4}$. 
The Pauli paramagnetic susceptibility and metallic behaviors of specific 
heat and electrical resistivity have been observed. The density of states at
the Fermi level $D(E_{\rm F})$ deduced 
from the results is about three times smaller than that of Sr$_{2}$RuO$_{4}$. 
Any indication of superconductivity intrinsic to Sr$_{2}$MoO$_{4}$ has not 
been 
observed down to 25\,mK, which may correspond to the smaller 
$D(E_{\rm F})$.
We discuss the origin of the difference in electronic states between 
Sr$_{2}$MoO$_{4}$ 
and Sr$_{2}$RuO$_{4}$.}
\kword{molybdate, Sr$_{2}$RuO$_{4}$, strongly-correlated electron system}
\begin{document}
\sloppy
\maketitle
The manifestation of spin-triplet superconductivity (SC) in the 
quasi-two-dimensional Fermi liquid state of Sr$_{2}$RuO$_{4}$ has 
marked an epoch in the study of SC in strongly-correlated 
electron systems.~\cite{ishida} 
One might expect that ferromagnetic spin fluctuations are
 essential for the origin 
of the spin-triplet SC, as observed in superfluid (p-wave) $^{3}$He. 
Rice and Sigrist
suggested the p-wave pairing in Sr$_{2}$RuO$_{4}$ immediately after the discovery of 
SC.~\cite{rice}
Their idea was based on the existence of ferromagnetism in 
SrRuO$_{3}$,~\cite{maeno1,yoshimura} 
the three-dimensional (3D) analogue to Sr$_{2}$RuO$_{4}$, and similar values of 
the Landau 
parameters to those of Fermi liquid $^{3}$He.

Concerning the electronic configuration in Sr$_{2}$RuO$_{4}$,
 the formal valence number of Ru ion 
 is 4$+$ with four 4d electrons in three $t_{2g}$ orbitals 
(4d$_{xy}$, 4d$_{yz}$, 4d$_{zx}$). The strong crystalline field for 4d
 electrons causes the Ru ion to have a low spin state with spin degree
 of freedom S$=$1,
 which means triple degeneracy and ferromagnetic correlations due to the Hund 
coupling. The normal state 
properties of 
Sr$_{2}$RuO$_{4}$ are well described quantitatively by the Fermi surface parameters
obtained from the quantum oscillation measurements.~\cite{andy} 
They confirmed
 that two cylindrical 
sheets of 4d$_{yz}$ and 4d$_{zx}$ orbitals and a cylindrical 
sheet of 4d$_{xy}$ orbitals consist of four 4d electrons of Ru$^{4+}$, which 
had been derived from the band-structure calculation.~\cite{oguchi}

A recent inelastic neutron scattering measurement on Sr$_{2}$RuO$_{4}$ has shown 
that
incommensurate peaks of dynamical magnetic susceptibility $\chi 
(\mib{Q})$ 
with $\mib{q}_0=(\pm 0.6\pi/a, \pm 0.6\pi/a, 0)$ exist instead of 
a ferromagnetic peak ($\mib{Q}\approx 0$).~\cite{sidis} The incommensurate
 peaks are ascribed to nesting vectors between Fermi surfaces with one-dimensional 
(1D) 4d$_{yz}$ and 4d$_{zx}$ orbitals.~\cite{mazin} This provides a new
physical aspect, the relation between the spin-triplet SC and the incommensurate 
 spin fluctuations.
If the SC 
mainly corresponds to this incommensurate spin fluctuation, 
the 1D 4d$_{yz}$ and 4d$_{zx}$ orbitals may play a vital role
 in the spin-triplet SC of 
Sr$_{2}$RuO$_{4}$. On the other hand, if the incommensurate spin fluctuation 
contributes negligibly
to the spin-triplet SC, the 2D 4d$_{xy}$ orbital is more important. Since the 
partial density of states for the 4d$_{xy}$ orbital is the 
highest among three t$_{2g}$ orbitals,~\cite{andy} most theoretical studies on
 the spin-triplet SC are based on this 2D 4d$_{xy}$ orbital.~\cite{ogata} 
 Recently, Takimoto has indicated 
the possibility of spin-triplet pairing symmetry in 
Sr$_{2}$RuO$_{4}$~\cite{takimoto} 
originating from the orbital fluctuations together with the incommensurate spin 
fluctuation 
observed 
by the inelastic neutron scattering.~\cite{sidis} The orbital 
fluctuations are supposed to be driven 
by the on-site Coulomb interaction between electrons in different 4d-t$_{2g}$
 orbitals, which are almost degenerate. At present, the important problem 
is whether or not the incommensurate spin fluctuation can induce the 
spin-triplet pairing. 
In addition, the role of the orbital degeneracy must be understood to
clarify the mechanism of the spin-triplet SC in Sr$_{2}$RuO$_{4}$.
 
An alternative route to understand the spin-triplet SC is provided by studying
other compounds
related to Sr$_{2}$RuO$_{4}$. Several metallic ruthenates were expected to reveal 
SC and were intensively examined.~\cite{ikeda,nakatsuji,nakatsuji2,cao}
 However, such ruthenates 
have not shown SC, unlike cuprates, where many charge-transfer-type
insulators can be made superconductive by ionic substitutions.
Hence, it is intriguing to find a ruthenium-free electronic analogue to 
Sr$_{2}$RuO$_{4}$. In order to provide a new system to be compared with 
Sr$_{2}$RuO$_{4}$, we have
focused on Sr$_{2}$MoO$_{4}$, because it is isostructural to Sr$_{2}$RuO$_{4}$
and has a similar electronic configuration. 
Mo ions in Sr$_{2}$MoO$_{4}$ formally possess the valence 4$+$ and have two 4d
 electrons in three $t_{2g}$ orbitals (4d$_{xy}$, 4d$_{yz}$, 4d$_{zx}$)
 with
 an equivalent orbital degeneracy (S$=$1) to Sr$_{2}$RuO$_{4}$. 
Moreover, Sr$_{2}$MoO$_{4}$ is expected to have three Fermi surfaces 
with the same topology as those of Sr$_{2}$RuO$_{4}$.~\cite{hase} These Fermi 
surfaces for the 1D orbitals with smaller Fermi wave vectors than those of 
Sr$_{2}$RuO$_{4}$ may also 
produce incommensurate spin fluctuations at $\mib{q}$ closer to the ferromagntic
wave vector ($\mib{Q}\approx 0$). Therefore, 
Sr$_{2}$MoO$_{4}$ is a good candidate not only for a
new superconductor but also for a counterpart to clarify the mechanism of
the spin-triplet SC in Sr$_{2}$RuO$_{4}$. 
In this Letter, we report the successful synthesis of polycrystalline 
Sr$_{2}$MoO$_{4}$ 
and the results of magnetic susceptibility $\chi$, electrical resistivity $\rho$ 
and specific heat $C$. We have no evidence for SC ascribable to Sr$_{2}$MoO$_{4}$ 
down to 25\,mK in the present results.

As for the synthesis of Sr$_{2}$MoO$_{4}$, there was a controversy over whether 
or not the 
phase actually existed. To our knowledge, Balz and Plieth first reported 
on Sr$_{2}$MoO$_{4}$,~\cite{balz} but McCarthy and Gooden could not obtain 
any Sr$_{2}$MoO$_{4}$ phase in their research.~\cite{mccarthy} 
 Lindblom and Rosen reported successful synthesis and 
deduced the lattice parameters by assuming I4/mmm 
symmetry.~\cite{lindblom} Very recently, Steiner and Reichelt performed 
X-ray Rietveld
analysis on a multi-phase sample including Sr$_{2}$MoO$_{4}$.~\cite{steiner}
They concluded that the crystal structure is the K$_{2}$NiF$_{4}$ type (Fig.~1),
which is the same as that of Sr$_{2}$RuO$_{4}$.~\cite{muller}
Nevertheless, systematic investigations on physical properties of 
Sr$_{2}$MoO$_{4}$ as well as definitive structural refinements have not yet been 
carried out since the synthesis of single-phase samples is extremely difficult.

The procedure for the sample synthesis is explained below. 
First, we prepared polycrystalline Sr$_{3}$MoO$_{6}$
from mixed raw materials of SrCO$_{3}$ (99.99\%) and MoO$_{3}$ (99.99\%) 
by firing in an Ar flow atmosphere at 1273\,K for 24 h with 
intermediate grindings. Then, Sr$_{3}$MoO$_{6}$ and Mo metal powder (99.9\%) 
were ground together
in a dry N$_{2}$ atmosphere so that the total composition was Sr$_{2}$MoO$_{4}$.
 The resultant mixture was pressed into pellets and then
sintered in a quartz tube under controlled
oxygen partial pressure at 1273\,K to 1473\,K. Details of the synthesis will be 
explained elsewhere.~\cite{shirakawa,shirakawa1}

 We characterized the obtained samples at room temperature by
 powder X-ray diffraction with Cu-K$\alpha$ radiation. DC magnetization above 
2\,K 
was measured using a SQUID magnetometer (Quantum Design, MPMS) and 
that between 25\,mK and 2\,K was measured using a SQUID magnetometer (SHE)
installed in 
a dilution refrigerator.~\cite{ootuka} Data below 2\,K were collected by 
moving the sample through the pickup coil under a field of 0.1\,Oe. 
Electrical resistivity $\rho (T)$ was
 measured by the ac method from 80\,mK to 300\,K. The samples for
 $\rho (T)$ and magnetization below 1\,K were covered 
with epoxy resin in order to avoid exposure to air. 
 Specific heat measurement between 0.7\,K and 60\,K were performed by the 
quasi-adiabatic heat-pulse method.
\begin{figure}
\leavevmode
\begin{center}
\epsfxsize=55mm
\epsfbox{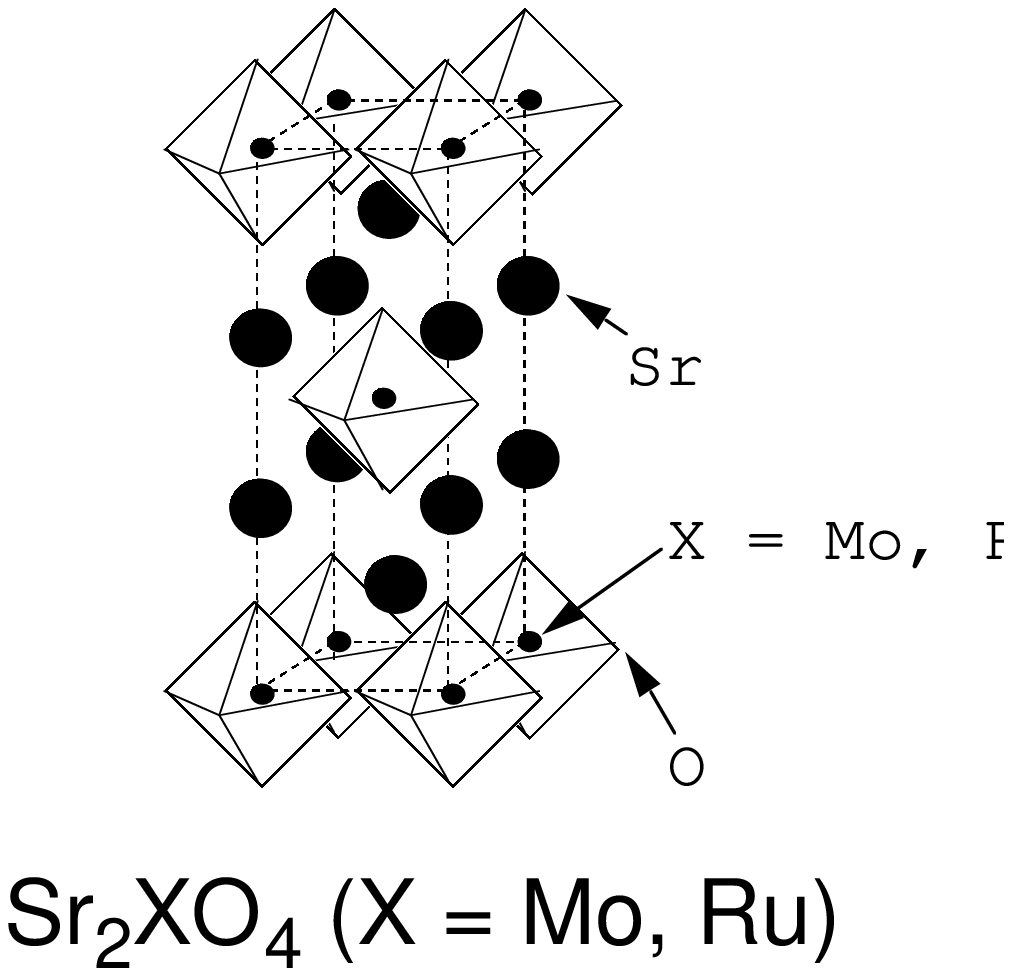}
\end{center}
\caption{Crystal structure (K$_{2}$NiF$_{4}$ type) of Sr$_{2}$RuO$_{4}$ and
 Sr$_{2}$MoO$_{4}$. The space group of the structure is I4/mmm.}
\label{fig:1}
\end{figure}
\begin{figure}
\leavevmode
\begin{center}
\epsfxsize=55mm
\epsfbox{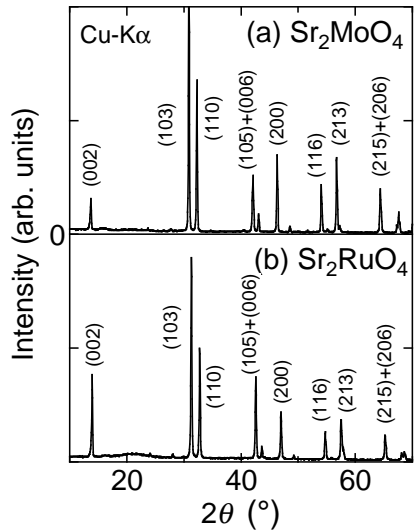}
\end{center}
\caption{Powder X-ray diffraction patterns of (a) Sr$_{2}$MoO$_{4}$ along with 
 (b) that of Sr$_{2}$RuO$_{4}$. Indices are based on the symmetry of I4/mmm.}
\label{fig:2}
\end{figure}
In Fig.~2, we show the powder X-ray pattern of Sr$_{2}$MoO$_{4}$ along with that for
a single phase of Sr$_{2}$RuO$_{4}$. The pattern of Sr$_{2}$MoO$_{4}$ without unidentified
peaks indicates that the obtained polycrystal is single-phased, except 
for a very weak peak of Mo metal. 
Moreover, these patterns suggest that both compounds share a common 
crystal structure. Indeed, the structural symmetry of Sr$_{2}$MoO$_{4}$ has been 
concluded to be I4/mmm from the electron diffraction analysis by Shirakawa 
 {\it et al.}~\cite{shirakawa1} The calculated lattice parameters 
of tetragonal Sr$_{2}$MoO$_{4}$ for the powder X-ray pattern
are $a=3.9168(4)$\,{\AA} and $c= 12.859(2)$\,{\AA} and those of 
Sr$_{2}$RuO$_{4}$ are
 $a=3.87073(2)$\,{\AA} and $c= 12.7397(1)$\,{\AA}.~\cite{chmaissem} The 
values of Sr$_{2}$MoO$_{4}$ are in 
good agreement with those of previous work.~\cite{steiner} Both $a$ and $c$
parameters are larger than 
those of Sr$_{2}$RuO$_{4}$ by about 1\%, owing to
the larger ionic radius of Mo$^{4+}$ than Ru$^{4+}$. We should note that 
the current samples still contain very small amounts of Mo metal and Mo$_{2}$C
 because the condition of the synthesis
 is highly subtle.~\cite{shirakawa1} Nevertheless, we could synthesize the 
samples with the amount of those phases less than 1\%.
\begin{figure}
\leavevmode
\begin{center}
\epsfxsize=55mm
\epsfbox{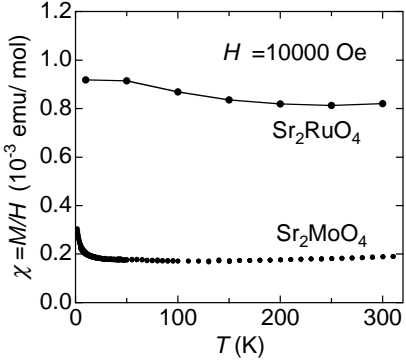}
\end{center}\caption{Magnetic susceptibility of Sr$_{2}$MoO$_{4}$ and Sr$_{2}$RuO$_{4}$ 
(ref.~25) before subtracting the Landau and the core diamagnetic
 contributions. The data of Sr$_{2}$RuO$_{4}$ was derived from the susceptibility 
of single crystals by calculating $(2\chi_{(H/\!\!/ab)}+\chi_{(H/\!\!/c)})/3$.}
\label{fig:3}
\end{figure}
\begin{figure}
\leavevmode
\begin{center}
\epsfxsize=55mm
\epsfbox{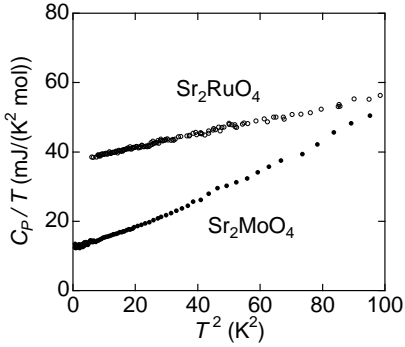}
\end{center}\caption{Temperature dependence of specific heat of Sr$_{2}$MoO$_{4}$ and 
 Sr$_{2}$RuO$_{4}$ (ref.~25). Data of $C/T$ are shown against $T^{2}$.}
\label{fig:4}
\end{figure}
Temperature dependence of magnetic susceptibility $\chi(T)=M/H$ is shown 
in Fig.~3 for both Sr$_{2}$MoO$_{4}$ and Sr$_{2}$RuO$_{4}$~\cite{maeno2}
 between 2\,K and
 300\,K.
 Sr$_{2}$MoO$_{4}$ reveals enhanced Pauli paramagnetic 
susceptibility, which is almost temperature independent except for 
the Curie-like increase with lowering
temperature below 20\,K. This probably corresponds to the existence of 
a small amount of magnetic impurities. Such Curie-like behavior is also 
observed in CaVO$_{3}$.~\cite{shirakawa2} 
The difference in the absolute value between Sr$_{2}$MoO$_{4}$ and Sr$_{2}$RuO$_{4}$
will be discussed later together with that of the electronic specific heat 
coefficient. No hysteresis is observed between zero-field-cooling (ZFC) and
field-cooling (FC) sequences. Hence, a ferromagnetically 
ordering component or a spin-glass state does not exist. In addition, there is no 
indication 
of a long-range magnetic ordering. 

The temperature dependence of specific heat $C_{P}(T)$ for Sr$_{2}$MoO$_{4}$ 
 is ordinary metallic behavior below 10\,K with a relatively large Sommerfeld
 coefficient $\gamma$, as shown in Fig.~4. The temperature dependence of 
$C_{P}(T)$ almost follows the relation $C_{P}(T)=\gamma T + \beta T^{3}$. The obtained electronic specific heat 
coefficient $\gamma$ is 12\,mJ/(K$^{2}$ Mo mol) for Sr$_{2}$MoO$_{4}$.
 Considering its Pauli paramagnetic 
susceptibility and normal metal behavior of $C_{P}(T)$, the ground state of 
Sr$_{2}$MoO$_{4}$ is probably dominated by renormalized quasi-particles, as 
observed in the Fermi liquid state. 
We did not observe the relevant anomaly in $C_{P}(T)$ below 10\,K to the
 Curie-type increase in 
$\chi(T)$. This is also consistent with the fact that the increase in $\chi(T)$ 
is due to a tiny amount of the magnetic impurity.

For Sr$_{2}$MoO$_{4}$, we try to evaluate the Wilson ratio, a dimensionless 
parameter concerned with correlation among electrons. It is necessary to 
determine the precise contributions of the Landau diamagnetic susceptibility 
especially 
when we have small mass enhancement of electrons. Hase has
 calculated the band structure for Sr$_{2}$MoO$_{4}$~\cite{hase} based on the atomic 
coordinates obtained by the prior X-ray Rietveld
 analysis.~\cite{steiner} According to their results, the 
density of states
$D(E_{\rm F})$ and electronic specific heat coefficient $\gamma_{calc}$ are 
2.1 states/eV and 5\,mJ/(K$^{2}$ Mo mol), respectively.~\cite{hase} The ratio of 
$\gamma$/$\gamma_{calc}$
should be nearly equal to the mass enhancement $m^{\ast}/m$, thus $m^{\ast}/m 
\approx \gamma/ 
\gamma_{calc} \approx 2$. The 
reduction factor by the Landau diamagnetism $1-(1/3)(m/m^{\ast})^{2}$ is about 
0.92 in the present case. 
Considering this diamagnetism and core diamagnetic susceptibility, we obtain the Pauli contribution 
$\chi_{\rm Pauli} = (\chi_{\rm obs}-\chi_{\rm core})/0.92 \approx 
3.2 \times 10^{-4} $\,emu/mol, where $\chi_{\rm obs}$ is the observed 
susceptibility and $\chi_{\rm core} = - 0.95 \times 10^{-4} $\,emu/mol.~\cite{selwood}
This value of $\chi_{\rm Pauli}$ is three times smaller than that of
 Sr$_{2}$RuO$_{4}$.~\cite{maeno2}

We obtained the Wilson ratio $R_{\rm W}=7.3 \times
 10^{4} \times \{\chi ({\rm emu/mol})\}\slash \{\gamma ({\rm mJ}/({\rm K}^{2}\,{\rm mol})\} = 1.9$, using the above value of $\chi_{\rm Pauli}$. This ratio is almost 
the 
same as that of Sr$_{2}$RuO$_{4}$.~\cite{maeno2}
The important point is that
 the value of $R_{\rm W}$ is considerably larger than unity, which is the value 
for
 the non-interacting free-electron system. This reflects the existence of 
substantial electron-electron correlation in Sr$_{2}$MoO$_{4}$, although it 
possesses smaller $D(E_{\rm F})$ than that of Sr$_{2}$RuO$_{4}$.

The electrical resistivity $\rho(T)$ of Sr$_{2}$MoO$_{4}$ presented 
in Fig.~5 shows metallic (d$\rho$/d$T \ge$0) behavior for the entire temperature 
range (80\,mK $\le T \le$ 300\,K). The ground state is supposed to be
 Fermi liquid, as recognized in Sr$_{2}$RuO$_{4}$,~\cite{maeno2} 
because of the temperature dependence of $C_{P}(T)$ and the Pauli paramagnetic 
$\chi(T)$. On the basis of the data of $ \rho(T)$ for polycrystals, we cannot 
draw any conclusions about its absolute value and the precise temperature 
dependence because of grain-boundary resistance and the mixing of both 
in-plane and out-of plane behaviors. It would be necessary 
to grow single-crystalline Sr$_{2}$MoO$_{4}$ to confirm whether the ground state 
is the Fermi liquid by observing the $T^{2}$-law in resistivity. 

We have paid a great deal of attention to the behavior of 
Sr$_{2}$MoO$_{4}$ below 4\,K to observe SC. In 
Fig.~6, we show $\rho(T)$ and $\chi(T)$ below 1\,K. Two drops of $\rho(T)$
at 0.9\,K and 0.4\,K are evident. The change at 0.9\,K is also present in 
the $\chi(T)$ 
data with a superconducting response. The volume fraction of this signal is, 
at most, about 1\%. We have concluded that this superconducting signal 
is in accordance with the existence of Mo metal as a trace amount of impurity in 
the Sr$_{2}$MoO$_{4}$ sample.
On the other hand, the other drop at 0.4\,K in $\rho(T)$ does not accompany 
\begin{figure}
\leavevmode
\begin{center}
\epsfxsize=55mm
\epsfbox{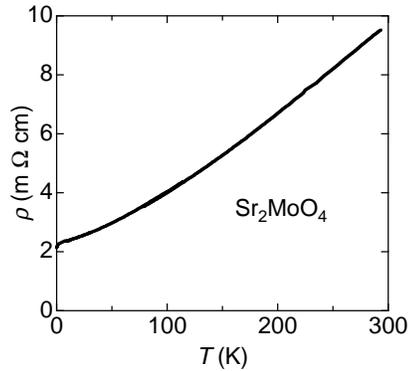}
\end{center}\caption{Temperature dependence of electrical resistivity of Sr$_{2}$MoO$_{4}$.}
\label{fig:5}
\end{figure}
\begin{figure}
\leavevmode
\begin{center}
\epsfxsize=55mm
\epsfbox{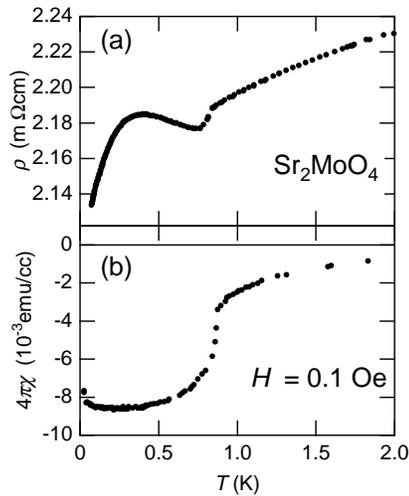}
\end{center}\caption{(a) Electrical resistivity $\rho(T)$ below 1\,K. (b) Magnetic 
susceptibility $\chi (T)$ under a field of 0.1\,Oe below 1\,K.}
\label{fig:6}
\end{figure}
any anomaly in $\chi(T)$. The negative value of $\chi(T)$ above 
0.9\,K is probably 
due to superconducting molybdenum carbides with 
$T_{\rm c}=3\mbox{--}4$\,K,~\cite{shirakawa} which was also found in $\chi(T)$ 
measured by the MPMS SQUID magnetometer.
 These results indicate that polycrystalline Sr$_{2}$MoO$_{4}$ is not 
superconducting above 25\,mK. 

Now we suggest the reasons why the polycrystalline Sr$_{2}$MoO$_{4}$ we 
obtained does not exhibit 
SC. The substantial difference between Sr$_{2}$MoO$_{4}$ 
and Sr$_{2}$RuO$_{4}$ is 
the density of states $D(E_{\rm F})$. The absolute values of 
$\chi(T)$ and $C_{P}(T)$, proportional to $D(E_{\rm F})$, of 
Sr$_{2}$MoO$_{4}$ are both about three times smaller than those of 
Sr$_{2}$RuO$_{4}$.
 This is nearly consistent with the preliminary results of band-structure 
 calculations for Sr$_{2}$MoO$_{4}$.~\cite{hase}

The origin of the difference of $D(E_{\rm F})$ between Sr$_{2}$MoO$_{4}$ 
and Sr$_{2}$RuO$_{4}$ is proposed below.
The ratio $d_{c}/d_{a}$, where $d_{c}$ 
($d_{a}$) is the distance between the transition-metal ion and the neighboring 
oxygen ion along the $c$-axis
 ($a$-axis), plays an essential role in energy-level splitting of the three 
$t_{2g}$ orbitals. According to Steiner and Reichelt,~\cite{steiner} the MoO$_{6}$ 
octahedron is more elongated along the $c$-axis than the RuO$_{6}$ octahedron in 
Sr$_{2}$RuO$_{4}$.
The ratio $d_{c}/d_{a}$ for Sr$_{2}$MoO$_{4}$ is 1.13~\cite{steiner} while 
$d_{c}/d_{a}$ for Sr$_{2}$RuO$_{4}$ is 1.07.~\cite{chmaissem} The larger 
$d_{c}/d_{a}$ will lead 
to the decreased population of 4d 
electrons in the 2D-4d$_{xy}$ orbital ($\gamma$ Fermi surface
 described in ref.~5). In other words, more dispersion along the 
$c$-axis is introduced 
and smaller $D(E_{\rm F})$ is expected. This smaller $D(E_{\rm F})$ is 
probably the main reason why Sr$_{2}$MoO$_{4}$ is not superconductive. 
Another important feature of strontium molybdates is paramagnetism in 
SrMoO$_{3}$ with a similar value of $\chi$ to that of 
Sr$_{2}$MoO$_{4}$.~\cite{bouchard,ikeda2} This shows a great contrast to the 
ferromagnetism in SrRuO$_{3}$,
which satisfies the Stoner criterion with the largest $D(E_{\rm F})$ 
in its paramagnetic state among ruthenates.~\cite{santi} The absence of 
significant ferromagnetic
spin fluctuation in SrMoO$_{3}$ may be related to the absence 
of SC in Sr$_{2}$MoO$_{4}$.
Despite our argument over the smaller $D(E_{\rm F})$
 above, further investigations on Sr$_{2}$MoO$_{4}$, such as single-crystal 
studies are highly anticipated, since Sr$_{2}$MoO$_{4}$ is the best system
to verify the importance of the orbital degeneracy or multiple Fermi surfaces 
to the spin-triplet SC in Sr$_{2}$RuO$_{4}$.

In summary, we have succeeded in synthesizing single-phase polycrystalline 
Sr$_{2}$MoO$_{4}$.
The first systematic studies of magnetic susceptibility, electrical resistivity and 
specific heat measurements 
have been performed down to low temperatures around $10^{-2}$\,K. We did
not observe any signs of superconductivity intrinsic to Sr$_{2}$MoO$_{4}$.
The value of the Wilson ratio shows remarkable electron-electron correlations.
 
The authors are grateful to I. Hase, T. Yanagisawa, T. Takimoto, Y. Yamaguchi,
C.~H. Lee and M. Koyanagi for their helpful supports and fruitful advice.


\begin{thebibliography}{99}
\bibitem{ishida} K. Ishida, H. Mukuda, Y. Kitaoka, K. Asayama, Z.~Q. Mao, 
Y. Mori and Y. Maeno: Nature (London) {\bf 396} (1998) 658.
\bibitem{rice} T.~M. Rice and M. Sigrist: J. Phys.\ Condens.\ Matter {\bf 7}
(1995) L643.
\bibitem{maeno1} Y. Maeno, S. Nakatsuji and S. Ikeda: Materials Science 
and Engineering B {\bf 63} (1999) 70.
\bibitem{yoshimura} K. Yoshimura, T. Imai, T. Kiyama, K.~R. Thurber, A.~W. 
Hunt and K. Kosuge: Phys.\ Rev.\ Lett.\ {\bf 83} (1999) 4397.
\bibitem{andy} A.~P. Mackenzie, S.~R. Julian, A.~J. Diver, G.~J. McMullan, M.~P. Ray, G.~G. Lonzarich, Y. Maeno, S. Nishizaki and T. Fujita: 
 Phys.\ Rev.\ Lett.\ {\bf 76} (1996) 3786.
\bibitem{oguchi} T. Oguchi: Phys.\ Rev.\ B {\bf 51} (1995) 1385.
\bibitem{sidis} Y. Sidis, M. Braden, P. Bourges, B. Hennion, S. NishiZaki,
Y. Maeno and Y. Mori: Phys.\ Rev.\ Lett.\ {\bf 83} (1999) 3320.
\bibitem{mazin} I.~I. Mazin and D.~J. Singh: Phys.\ Rev.\ Lett.\ {\bf 82} (1999) 
4324.
\bibitem{ogata} T. Kuwabara and M. Ogata:
to be published in Phys.\ Rev.\ Lett.\
\bibitem{takimoto} T. Takimoto: unpublished.
\bibitem{ikeda} S.~I. Ikeda, Y. Maeno, S. Nakatsuji, M. Kosaka and Y. Uwatoko:
Phys.\ Rev.\ B {\bf 62} (2000) R6089.
\bibitem{nakatsuji} S. Nakatsuji, S. Ikeda and Y. Maeno: J. Phys.\ Soc.\ Jpn.\
{\bf 66} (1997) 1868.
\bibitem{nakatsuji2} S. Nakatsuji and Y. Maeno: Phys.\ Rev.\ Lett.\
{\bf 84} (2000) 2666.
\bibitem{cao} G. Cao, S. McCall, J.~E. Crow and R.~P. Guertin:
Phys.\ Rev.\ Lett.\ {\bf 78} (1997) 1751.
\bibitem{hase} I. Hase: unpublished.
\bibitem{balz} V.~D. Balz and K. Plieth: Z. Elektrochem.\ {\bf 59}
 (1955) 545. 
\bibitem{mccarthy} G.~J. McCarthy and C.~E. Gooden: J. Inorg.\ Nucl.\ Chem.\
 {\bf 35} (1973) 2669.
\bibitem{lindblom} B. Lindblom and E. Rosen: Acta Chemica Scandinavica
 A {\bf 40} (1986) 452.
\bibitem{steiner} U. Steiner and W. Reichelt: Z. Naturforsch.\ {\bf 53 b}
 (1998) 110.
\bibitem{muller} Hk. M\"uller-Buschbaum and J. Wilkens:
Z. Anorg.\ Allg.\ Chem.\ {\bf 591} (1990) 161.
\bibitem{shirakawa} N. Shirakawa and S.~I. Ikeda: to be published in Physica 
C.
\bibitem{shirakawa1} N. Shirakawa, S.~I. Ikeda and H. Matsuhata: unpublished.
\bibitem{ootuka} Y. Ootuka and N. Matsunaga: J. Phys.\ Soc.\ Jpn.\ {\bf 59}
 (1990) 1801.
\bibitem{chmaissem} O. Chmaissem, J.~D. Jorgensen, H. Shaked, S. Ikeda
and Y. Maeno: Phys.\ Rev.\ B {\bf 57} (1998) 5067.
\bibitem{maeno2} Y. Maeno, K. Yoshida, H. Hashimoto, S. Nishizaki, 
S. Ikeda, M. Nohara, T. Fujita, A.~P. Mackenzie, N.~E. Hussey, 
J.~G. Bednorz and F. Lichtenberg: J. Phys.\ Soc.\ Jpn.\ {\bf 66} (1997) 1405.
\bibitem{shirakawa2} N. Shirakawa, K. Murata, H. Makino, F. Iga and Y. Nishihara:
J. Phys.\ Soc.\ Jpn.\ {\bf 64} (1995) 4824.
\bibitem{selwood} P.~W. Selwood: Magnetochemistry (Interscience 
Publishers, New York, 1956) 2nd ed., p.~78. 
\bibitem{bouchard} G.~H. Bouchard,~Jr. and M.~J. Sienko: Inorg.\ Chem.\
 {\bf 7} (1968) 441.
\bibitem{ikeda2} S.~I. Ikeda and N. Shirakawa: to be published in Physica C.
\bibitem{santi} G. Santi and T. Jarlborg: J. Phys.: Condens.\ Matter 
 {\bf 9} (1997) 9563.
\end{thebibliography}
\end{document}